\documentclass[letterpaper, 11pt]{article}

\usepackage[margin=0.75in]{geometry}
\usepackage{amsfonts,
	amsmath,
	graphicx,
	subfigure,
	url}
\usepackage{tikz}
\usepackage{subfigure}
\usepackage[mathscr]{euscript}
\usepackage[colorlinks=true, bookmarksopen,
linkcolor={blue},
anchorcolor={black},
citecolor={red},
filecolor={magenta},
menucolor={black},
plainpages=false,pdfpagelabels,
urlcolor={blue}]{hyperref}

\title{Improving Runtime Performance of Tensor Computations\\ using Rust From Python}

\author{
	Kimmie Harding\thanks{New Jersey Institute of Technology (\texttt{kah46@njit.edu}); Sandia National Laboratories (\texttt{kahardi@sandia.gov})} 
	\and 
	Daniel M. Dunlavy\thanks{Sandia National Laboratories (\texttt{dmdunla@sandia.gov})}
}

\date{}

\begin{document}
	
	\maketitle
	
	\begin{abstract}
            In this work, we investigate improving the runtime performance of key computational kernels in the Python Tensor Toolbox (\texttt{pyttb}), a package for analyzing tensor data across a wide variety of applications. Recent runtime performance improvements have been demonstrated using Rust, a compiled language, from Python via extension modules leveraging the Python C API---e.g., web applications, data parsing, data validation, etc. Using this same approach, we study the runtime performance of key tensor kernels of increasing complexity, from simple kernels involving sums of products over data accessed through single and nested loops to more advanced tensor multiplication kernels that are key in low-rank tensor decomposition and tensor regression algorithms. In numerical experiments involving synthetically generated tensor data of various sizes and these tensor kernels, we demonstrate consistent improvements in runtime performance when using Rust from Python over 1) using Python alone, 2) using Python and the \texttt{Numba} just-in-time Python compiler (for loop-based kernels), and 3) using the \texttt{NumPy} Python package for scientific computing (for \texttt{pyttb} kernels).
	\end{abstract}
	
	\begin{figure}[b!]
		\centering
		\includegraphics[width=\textwidth]{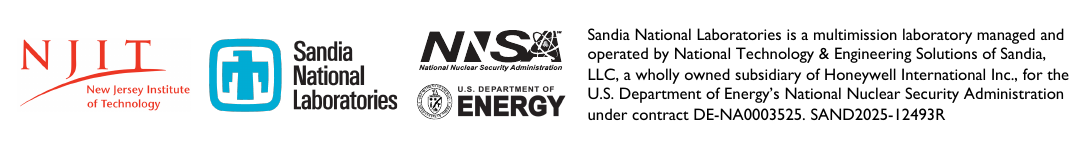}
	\end{figure}
	
		\section{Introduction}
\label{sec:intro}
Tensors, or multi-dimensional ($d$-way) arrays~\cite{BaKo2025}, can be used to represent complex relationships in a variety of data analysis applications, such as deep learning~\cite{PaKoChOl2024}, quantum computing~\cite{MaYa2022}, quantum chemistry~\cite{TaHuNiMa2025}, signal processing~\cite{TOKCAN2026110191}, neuroscience~\cite{PeStCa2026}, scientific computing~\cite{Kh2018}, and others.
Several key tensor algorithms and operations, such as the sparse tensor times vector~(TTV) product and sparse matricized tensor times Khatri-Rao product~(MTTKRP), require computation that scales exponentially with $d$. Therefore, improving the runtime performance of such computational kernels and other tensor operations is critical for effective tensor data analysis.

The Python Tensor Toolbox, \texttt{pyttb}~\cite{pyttb}, is a recently developed software package for tensor operations. In the work presented here, we explore the possibility of improving runtime performance of \texttt{pyttb} on CPUs by implementing key tensor kernels in the compiled language Rust~\cite{Rust_Book} that are then called from Python. Several strategies for improving runtime performance in Python include leveraging compiled-code extensions through the Python Foreign Function Interface (FFI)~\cite{10.1007/978-3-031-85902-1_4} and just-in-time (JIT) compilation~\cite{10.1145/2833157.2833162}, and we compare several such approaches in the numerical experiments presented below. We focus on using Rust from Python via the Python FFI, as this approach has shown improved performance for a variety of applications---e.g., web applications, databases, file I/O, data parsing/validation, etc.~\cite{Rao2023}. However, there is limited research available regarding using Rust from Python for numerical computing in general and tensor operations specifically. Although there exist many examples of improving runtime performance of Python software using C from Python, including previous work involving tensor algorithms~\cite{PeDu2014}, we focus primarily on using Rust from Python, as a detailed comparison using Rust from Python and using C from Python is beyond the scope of this work. 

We test the hypothesis that using Rust from Python~\cite{PyO3_Project_and_Contributors_PyO3} for tensor operations leads to faster runtime performance over using Python alone by implementing several tensor kernels of increasing complexity and comparing average runtimes over multiple trials of each approach.
Specifically, our contributions are as follows:
\begin{itemize}
	\item Demonstration of runtime performance improvements greater than 2 orders of magnitude using Rust from Python over Python for two simple tensor kernels: vector dot product and dense matrix-vector product; and
	\item Demonstration of runtime performance improvements by approximately 1 order of magnitude using Rust from Python over Python for a more advanced tensor kernel: sparse tensor times vector product.
\end{itemize}
The remainder of the paper is structured as follows: Section~\ref{sec:background} provides background material on tensor computations and related work; Section~\ref{sec:methods} describes the approach for comparing runtime performance of Python, Rust, and using Rust from Python for several tensor kernels; and  Section~\ref{sec:results} presents the numerical experiments. We then present our conclusions and ideas for potential future work in Section~\ref{sec:conc}.

\section{Background}
\label{sec:background}
In this section, we provide an overview of tensors and the \texttt{pyttb} software package, including the necessary background for using Rust from Python to achieve runtime performance improvements for tensor kernels. In addition, we discuss alternative approaches for potential runtime performance improvements of tensor kernels that could be considered in future research.

\subsection{Tensors, Python, and Rust}
\label{sec:background:tensors}
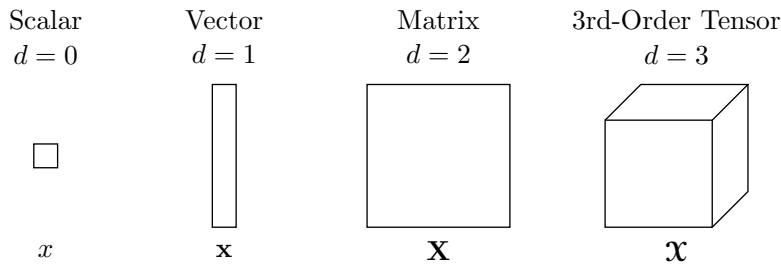
\begin{figure}[ht!]
	\begin{center}
		\resizebox{0.6\textwidth}{!}{
			\begin{tikzpicture}
				\node[align=center,scale=1.5] at (-3.75, 4) {Scalar \\ \( d = 0 \)};
				
				\draw[thick] (-4,1.25) rectangle (-3.5,1.75);
				
				\node[scale=1.5] at (-3.75, -0.5) {\( x \)};
				
				\node[align=center,scale=1.5] at (0, 4) {Vector \\ \( d = 1 \)};
				
				\draw[thick] (-0.25,0) rectangle (0.25,3);
				
				\node[scale=1.5] at (0, -0.5) {\( \mathbf{x} \)};
				
				\node[align=center,scale=1.5] at (4.5, 4) {Matrix \\ \( d = 2 \)};
				
				\draw[thick] (3,0) rectangle (6,3);
				
				\node[scale=1.5] at (4.5, -0.5) {\( \mathbf{X} \)};
				
				\node[align=center,scale=1.5] at (9.5, 4) {3rd-Order Tensor \\ \( d = 3 \)};
				
				\draw[thick] (8,0) -- (10.25,0) -- (10.25,2.25) -- (8,2.25) -- cycle; 
				\draw[thick] (8,2.25) -- (8.75,3) -- (11,3) -- (10.25,2.25); 
				\draw[thick] (11,3) -- (11,0.75) -- (10.25,0); 
				
				\node[scale=1.5] at (9.5, -0.5) {\( \boldsymbol{\mathscr{X}} \)};

			\end{tikzpicture}
		}
	\end{center}
	\caption{Illustration of tensors of various dimensions.}
	\label{fig:tensor}
\end{figure}

\emph{Tensors} refer to general data arrays with $d \geq 0$ dimensions. Thus, scalars, vectors, and matrices---as well as arrays with $d > 2$---are all examples of tensors, as illustrated in Figure~\ref{fig:tensor}.
Throughout this paper, we use standard tensor notation from~\cite{BaKo2025} as follows: scalars ($d=0$) are denoted by lowercase letters (e.g.~\( x\)), vectors ($d=1$) are denoted as bold lowercase letters (e.g.,~\( \mathbf x\)), matrices ($d=2$) are denoted as bold uppercase letters (e.g.,~\(\mathbf X\)), and general tensors ($d \geq 3$) are denoted as bold Euler script letters (e.g.,~\(\boldsymbol{\mathscr{X}}\)). All tensor elements in our current work are real-valued (i.e., taken from the set of real values, $\mathbb{R}$).
Furthermore, tensor elements are defined by their indices ranging from 1 to the size of their dimensions; for example, vector elements are denoted as $x_{i_1}$ for ${\mathbf x} \in \mathbb{R}^{n_1}$ with $i_1 \in \{1,\dots,n_1\}$; matrix elements are denoted as $x_{i_1,i_2}$ for  ${\mathbf X} \in \mathbb{R}^{n_1 \times n_2}$ with $i_1 \in \{1,\dots,n_1\}$ and $i_2 \in \{1,\dots,n_2\}$; and general tensor elements are denoted as $x_{i_1, i_2,\dots,i_d}$ for ${\boldsymbol{\mathscr{X}}} \in \mathbb{R}^{n_1 \times n_2 \times \cdots \times n_d}$ with $i_1 \in \{1,\dots,n_1\}$, $i_2 \in \{1,\dots,n_2\}$, and $i_d \in \{1,\dots,n_d\}$.

The Python Tensor Toolbox~(\texttt{pyttb})~\cite{pyttb} is a Python implementation of the highly cited Tensor Toolbox for MATLAB~(\texttt{TTB})~\cite{TTB_MATLAB}. Previous work indicated that the implementation of several tensor algorithms in Python can lead to faster runtimes over MATLAB~\cite{PeDu2014}, motivating the creation of \texttt{pyttb} to support tensor research and data analysis consistent with \texttt{TTB} for Python users. Currently, tensor data classes in \texttt{pyttb} are defined using the \texttt{ndarray} data structure from the \texttt{NumPy} Python package. Instances of \texttt{ndarray}s are stored in memory as contiguous blocks and most functions involving \texttt{ndarray}s utilize vectorized operations for improved performance~\cite{2020NumPy-Array}. Dense tensors are defined using an \texttt{ndarray} for the data values, and sparse tensors are defined using an \texttt{ndarray} for the indices of the nonzero values and an \texttt{ndarray} for the values corresponding to each of those indices. The \emph{shape}, or sizes of the dimensions of the tensor, is defined using a tuple for both dense and sparse tensors. These attributes of the \texttt{pyttb} implementation are important for understanding how we are using Rust from Python for the various tensor kernels we investigated in our work and will be discussed in more detail in Section~\ref{sec:methods}.

The Python language manages memory through reference counting and garbage collection, incurring overhead costs and potentially impacting runtime performance.
Reference counting is maintained through Python's Global Interpreter Lock (GIL), which only allows one thread to control the Python interpreter at a time\footnote{Although recent versions of Python provide support for an experimental feature of free-threading by releasing the GIL~\cite{freeThreadingPython}, which can leverage parallelism for runtime performance improvements, the use of this feature is beyond the scope of this work.}. In CPython, which we use exclusively in the research presented here, the Python C API~\cite{pythoncapi} handles the interaction between C (or C-compatible) code and the Python interpreter by managing data type conversions and reference counting across the FFI boundary~\cite{van1995python}. In Rust, memory is managed through \emph{ownership}, a set of rules checked at compile time to ensure memory safety~\cite{Rust_Book}. We use the \texttt{PyO3} Rust crate\footnote{Crates are packages external to the core Rust language that are available via Rust's Package Registry: \url{https://crates.io} .} to provide interoperability between the different memory managers in Rust and the Python C API~\cite{PyO3_Project_and_Contributors_PyO3}. Python-compatible code is created and packaged from the \texttt{PyO3}-enabled Rust code using the \texttt{Maturin}~\cite{maturin} Python package.


\subsection{Related Work}
\label{sec:background:related}
As mentioned above, the \texttt{NumPy} Python package provides vectorized operations of its data structures that utilize compiled C code, avoiding explicit loops/indexing and resulting in improved performance~\cite{2020NumPy-Array}. As some of these vectorized functions are already used in \texttt{pyttb} (via methods involving \texttt{ndarray}s), we can use current implementations of tensor kernels in \texttt{pyttb} as baselines for measuring runtime performance improvements, thus comparing using Rust from Python to using C from Python. Although, as noted above, the goal of our work presented here is not a comprehensive comparison of using Rust from Python and using C from Python, but rather if using Rust from Python is a viable option for improving runtime performance of tensor kernels originally implemented in Python.

The \texttt{Numba} Python package includes a just-in-time (JIT) compiler that optimizes Python code containing loops and \texttt{NumPy} data structures~\cite{10.1145/2833157.2833162}. \texttt{Numba} provides function decorators to denote Python functions to be compiled into machine code during the initial call to each function. This approach can reduce overall development cost and complexity but may not actually lead to improved performance if \texttt{Numba}'s JIT compiler cannot successfully compile a function into machine code as requested. For example, if the intermediate representation of variables defined in a function cannot be identified at compilation time, the function is compiled in \emph{object mode}, which has lower performance than when compiling fully into machine code using \emph{no python mode}~\cite{10.1145/2833157.2833162}. A detailed comparison between using \texttt{Numba} and using Rust from Python is beyond the scope of this work, but we include several examples of comparisons in the numerical experiments in Section~\ref{sec:results} to illustrate some of the performance improvements that can be gained using \texttt{Numba}.

As stated above in Section~\ref{sec:intro}, using Rust from Python for improved runtime performance has been demonstrated in various applications, including the \texttt{Polars} DataFrame package~\cite{Ped2023}, the \texttt{TikToken} natural language processing package~\cite{tiktoken}, and the \texttt{Pydantic} data validation package~\cite{Col2022}. However, there is little published evidence of runtime performance improvement associated with packages focused on numerical computing (and specifically on tensor operations), thus motivating our work presented here. 

\section{Methodology}
\label{sec:methods}
We test our hypothesis that using Rust from Python leads to faster runtime performance of tensor operations in \texttt{pyttb} using computational kernels of increasing complexity run over a range of various data sizes. The computational kernels are implemented in Python, Rust, and using Rust from Python, and the average runtimes over multiple trials are measured as a function of the size of the data. Specifically, we focus here on the following tensor kernels:
\begin{itemize}
	\item vector dot product (Section~\ref{sec:methods:dot});
	\item dense matrix-vector product  (Section~\ref{sec:methods:matvec}); and
	\item sparse tensor times vector product (Section~\ref{sec:methods:sttv}).
\end{itemize}

For each tensor kernel, we run an experiment consisting of $n_{trials}$ trials for each combination of kernel implementation and size of data to account for system runtime variability. All data arrays contain randomized 64-bit floating-point values (i.e., \texttt{numpy.float64} in Python and \texttt{f64} in Rust) uniformly sampled from $[0,1]$ to reflect the standard data types used in numerical computing in general and \texttt{pyttb} specifically. New data arrays are generated for each trial of the loop-based kernels: vector dot product and dense matrix-vector product. However due to generation costs, one data array is generated for all trials of the sparse tensor times vector product.  All the values are sampled using the \verb|uniform| function from the \texttt{random} module in the  \texttt{NumPy} Python package and using the \verb|Uniform| struct from the \texttt{rand\_distr} Rust crate. The data sizes chosen in our experiments were determined as those sizes that resulted in runtimes greater than milliseconds for the Python tensor kernel implementations to reduce the impact of noise in the timing instrumentation. 

We implement all tensor kernels using the \texttt{NumPy} Python package \texttt{ndarray} data structure. However, only the sparse tensor times vector product kernel in \texttt{pyttb} leverages some of the vectorized functions from the \texttt{NumPy} Python package. Similarly, the Rust implementations use data structures from the \texttt{Ndarray} Rust crate~\cite{ndarray_rust}, and only the sparse tensor times vector product kernel employs various \texttt{Ndarray} Rust crate methods for comparable functionality to the methods in the \texttt{NumPy} Python package.

The timing instrumentation in Python and Rust both use monotonic clocks that measure the duration between the two reference points. In Python, we use the \verb|perf_counter| function from the \texttt{time} module. In Rust, we use the \verb|now| and \verb|elapsed| methods of the \verb|Instant| struct from the \texttt{std} Rust crate. Thus, there could be some minor timing discrepancies when comparing runtimes between Rust and Python in our experiments. However, all timing results within Python (i.e., for the Python loop-based kernels, \texttt{NumPy}, \texttt{Numba}, and \texttt{pyttb}) are consistent across experiments. When using Rust from Python with the \texttt{PyO3} Rust crate, overhead costs are included in the runtimes of the initial call per Python script execution. The specific sources of the overhead cost for the first call are uncertain---possibilities include crossing the FFI boundary, loading the Python module, data type coversions, etc.---however, future research is required to determine the contribution of each source. To control for these overhead costs in our experiments, we compute each kernel once before measuring runtimes. Furthermore, we estimate this overhead when using Rust from Python for the vector dot product and the dense matrix-vector product at each of the data sizes by computing the difference in average runtimes of the first call and the average runtimes of the $n_{trials}$ calls after the first call for each experiment. Reports of these estimated costs are reported in the numerical results presented in Section~\ref{sec:results}.

\subsection{Vector Dot Product}
\label{sec:methods:dot}
The vector dot product computes the scalar product of two vectors (first-order tensors) as the summation of the products of the elements of each vector at the same indices.
The vectors must have the same number of elements in order to calculate the dot product.
The equation for scalar output is as follows:
\begin{align}
	\mathbf x \cdot \mathbf y = \sum_{{i_1}=1}^{n_1} x_{i_1} y_{i_1} = x_1 y_1 + x_2 y_2 + \cdots + x_{n_1} y_{n_1} \; .
	\label{eq:vdp}
\end{align}

We implement Equation~\ref{eq:vdp} using a single loop over the two vectors in Python, Rust, and using Rust from Python. We also compare the vector dot product using \texttt{Numba}'s JIT-compiled version of the Python implementation. When using Rust from Python and \texttt{Numba}, we estimated the costs of the first call of using Rust from Python and the \texttt{Numba} JIT compiler, respectively, to illustrate the additional overhead in using these approaches. Note, the overhead of the first call of using the \texttt{Numba} Python package includes both the JIT compile time in addition to the other overhead costs mentioned above.

\subsection {Dense Matrix-Vector Product}
\label{sec:methods:matvec}
The dense matrix-vector product computes an output vector whose elements consist of the vector dot product between the rows of a matrix (second-order tensor) and a vector (first-order tensor). Mathematically, a dense matrix-vector product of ${\mathbf A} \in \mathbb{R}^{n_1 \times n_2}$ and ${\mathbf x} \in \mathbb{R}^{n_2}$ is defined as follows:
\begin{align}
	\mathbf A \times \mathbf x = \left[ \sum_{{n_2}=1}^{n_2} a_{1{i_2}} x_{i_2}, \; \sum_{{i_2}=1}^{n_2} a_{2{i_2}} x_{i_2}, \; \dots, \; \sum_{{i_2}=1}^{n_2} a_{{n_1}{i_2}} x_{i_2}\right]^\top \; ,
	\label{eq:mvp}
\end{align}
where the result is a column vector of length $n_1$.
We implement Equation~\ref{eq:mvp} using a nested loop in Python, Rust, and using Rust from Python. The outer loop computes the elements of the resulting column vector (first-order tensor)---i.e., the elements of the right hand side of Equation~\ref{eq:mvp} above. The inner loop computes the sums in each of those elements of the resulting vector. 

We define multiple experiments involving the dense matrix-vector product corresponding to matrices with an 1) increasing number of rows ($n_2$ constant), 2) increasing number of columns ($n_1$ constant), and 3) increasing number rows and columns ($n_1$ and $n_2$ equal). Note that computing with second-order tensors requires the additional consideration of \emph{memory layout}, as the elements of matrices can be stored in memory in row-major or column-major order---defined in the \texttt{NumPy} Python package as ``\texttt{C}'' and ``\texttt{F}'' (denoting Fortran) order, respectively. All data structures in the current implementation of \texttt{pyttb} use column-major order (for consistency with \texttt{TTB}), whereas the corresponding data structures in the \texttt{Ndarray} Rust crate use row-major order, and these differences may impact the runtime performance comparisons. Further research could be conducted to assess whether \texttt{pyttb} could be improved by providing more flexible memory layout options. Although preliminary studies demonstrated that memory layout can impact runtime performance, a detailed assessment of the impact of memory layout in tensor operations is beyond the scope of the current work. As in the vector dot product experiment described above in Section~\ref{sec:methods:dot}, we also compare runtime performance using JIT-compiled versions of the Python implementation via \texttt{Numba}, and runtime costs associated with the first call of using Rust from Python and the \texttt{Numba} JIT compiler are estimated in a similar way.

\subsection{Sparse Tensor Times Vector Product}
\label{sec:methods:sttv}
The sparse tensor times vector (TTV) product multiples a $d$-way tensor with $n_{v}$ vectors resulting in a tensor with $d-n_v$ dimensions, where $n_{v} \in \{1,\dots,d\}$ and the size of each vector must correspond to the size of one of the dimensions of the $d$-way tensor~\cite{BaKo2025}. For example, for the TTV product of a sparse third-order tensor ($d=3$) and a single vector ($d=1$ and $n_v=1$), the result is a tensor of dimension $d-n_v = 3-1=2$ (i.e., a matrix). Mathematically, the TTV product of a $d$-way tensor ${\boldsymbol{\mathscr{A}}} \in \mathbb{R}^{n_1 \times n_2 \times \cdots \times n_d}$ and a vector ${\mathbf x} \in \mathbb{R}^{n_k}$ can be defined element-wise as follows:
\begin{align}
	\left(\boldsymbol{\mathscr{A}} \times_k \mathbf x\right)_{i_1,\dots,i_{k-1},i_{k+1},\dots,i_d} = \sum_{{i_k}=1}^{n_k} a_{i_1, i_2,\dots,i_d} x_{i_k} \; ,
	\label{eq:ttv}
\end{align}
where $\times_k$ denotes mode-$k$ tensor multiplication~\cite[Sec.~3.3]{BaKo2025} and $k \in \{1,\dots,d\}$.

Sparse tensors contain a large percentage of zero values; therefore, only the nonzero elements are stored to support efficient use of memory. This can be beneficial for data analysis, as loops over the sparse tensor instances only iterate over the indices associated with nonzero values. The sparse TTV serves as the foundation of the sparse matricized tensor times Khatri-Rao product (MTTKRP) kernel, the main computational kernel used in important tensor algorithms, such as low-rank tensor decompositions~\cite{BaKo2025} and tensor-on-tensor regression~\cite{Lock18}. Therefore, improving the runtime performance of the sparse TTV kernel in \texttt{pyttb} will also improve the runtime performance of the sparse MTTKRP and these important tensor algorithms.

For the experiment involving the TTV kernel, we use the sparse tensor function \texttt{sptensor.ttv} from \texttt{pyttb} (\texttt{pyttb.sptensor.ttv}) as the Python implementation. Although \texttt{pyttb} supports general dimensions for sparse tensors (e.g. $d=1$ or $d=2$) and the ability to compute against multiple vectors in one operation, this experiment only focuses on the case of a third-order ($d$=3) tensor times one vector as an illustration of typical runtime performance for the sparse TTV kernel.
Furthermore, this experiment isolates the runtime performance of the TTV kernel by removing all error checking from the \texttt{pyttb} code. Although \texttt{pyttb} converts the output to a dense tensor when the percentage of nonzero elements exceeds 50\%, this experiment removes this conversion step as well. 

The sparse TTV kernel in \texttt{pyttb} leverages several vectorized functions of the \texttt{ndarray} class in the \texttt{NumPy} Python package that are not implemented in the \texttt{Ndarray} Rust crate: \verb|setdiff1d|,  \verb|arrange|, \verb|unique|, \verb|accumarray|, \verb|flatten|, \verb|squeeze|, \verb|nonzero|. Therefore, our implementation in Rust does not correspond exactly line-by-line to the implementation in \texttt{pyttb}. Instead, our implementation provides equivalent functionality of the TTV kernel using a combination of available functions provided in the \texttt{Ndarray} Rust crate, Rust iterators, and several new functions created specifically for this experiment. Note that we made no attempt to improve the current sparse TTV implementation in \texttt{pyttb}, and future work could attempt to improve the \texttt{pyttb} implementation to provide a more equitable comparison. Furthermore, as no sparse tensor implementation currently exists in Rust, our implementation of the TTV kernel when using Rust from Python returns the individual arrays for \verb|newsubs|, \verb|newvals|, and \verb|newshape|, which are used to instantiate a new instance of \verb|pyttb.sptensor| after the return to Python (which is equivalent to the final step of the \texttt{pyttb.sptensor.ttv} implementation). This runtime cost of the \verb|sptensor| instantiation in \texttt{pyttb} is included in the average runtimes presented in the numerical results in Section~\ref{sec:results}.

The sparse TTV kernel is tested using a constant density of nonzero elements across the various data sizes investigated in the experiment.
The data tensors, $\boldsymbol{\mathscr{A}}$, are also instantiated with random values using the \verb|pyttb.sptenrand| method, which generates the nonzero elements by sampling uniformly from $[0,1]$. For consistency across the experiment, in all trials for all data sizes, the sparse TTV is computed using third-order tensors with dimensions of equal sizes, $n_1 = n_2 = n_3$, multiplied by a single vector (first-order tensor) of size $n_2$ along dimension $k=2$.

\section{Numerical Experiments}
\label{sec:results}
All numerical experiments were performed on a system with an AMD 1.8 GHz EPYC 9345P processor and 768GB RAM, running Red Hat Enterprise Linux Release (RHEL) 9.5 and the following versions of Python, Rust, and their respective packages and crates:

\begin{itemize}
	\item Python 3.12 and the following packages:
	\subitem \texttt{pyttb} 1.8.2, \texttt{NumPy} 2.2.6, \texttt{Maturin} 1.9.4, \texttt{Numba} 0.61.2; and
	\item Rust 1.81 and the following crates:
	\subitem \texttt{PyO3} 0.25.0, \texttt{Ndarray} 0.16.1, \texttt{rust-numpy} 0.25.0, \texttt{Rand} 0.9.2, \texttt{rand\_distr} 0.5.1.
\end{itemize}

For each tensor kernel, we computed the average runtimes across $n_{trials} = 30$ trials for each size of data. 

\subsection{Vector Dot Product}
\label{sec:results:dot}
\begin{figure}[ht!]
	\centering
	\includegraphics[height=0.275\textheight]{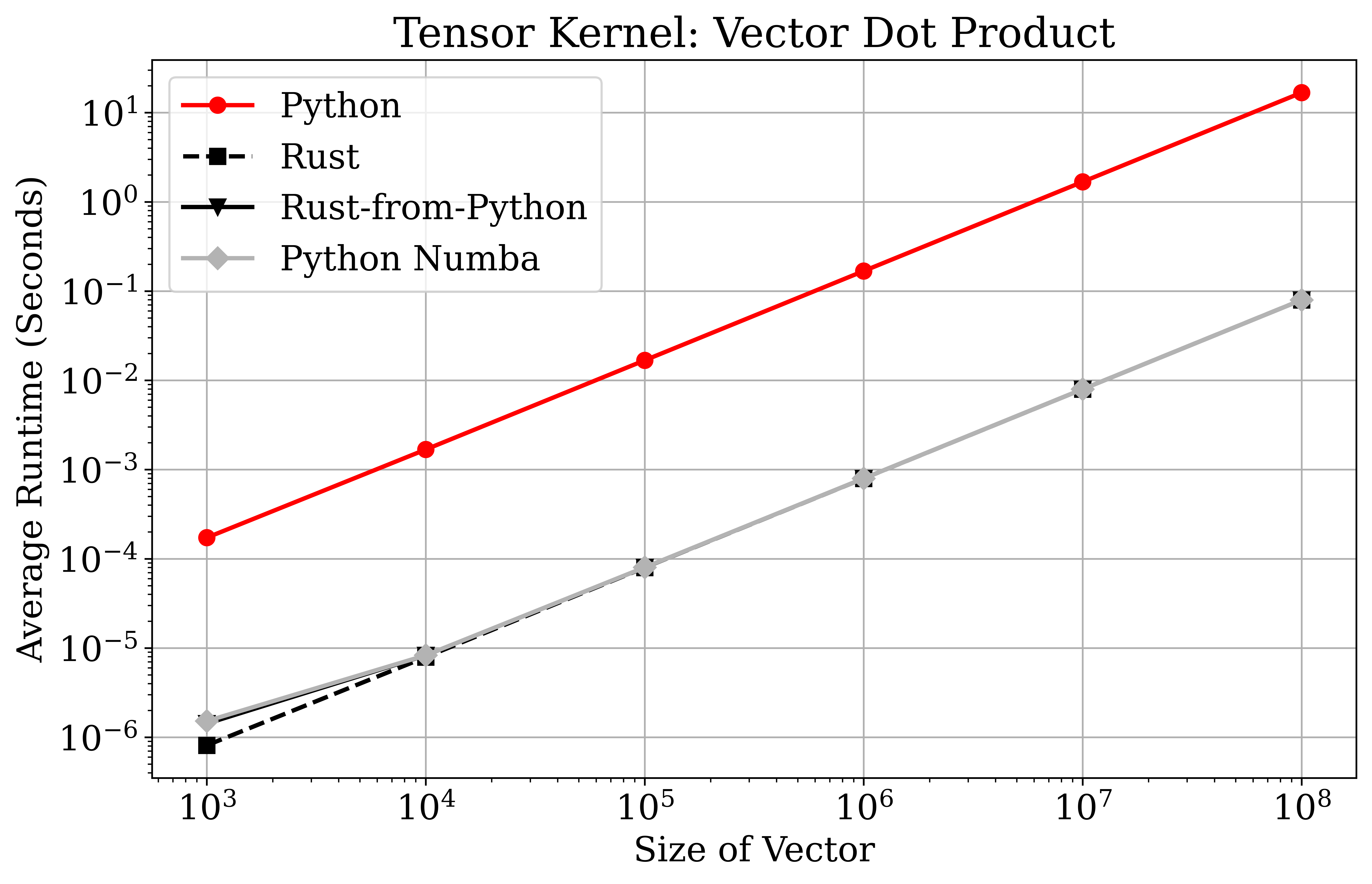}
	\caption{Average runtime of the vector dot product as a function of vector size.}
	\label{fig:dot}
\end{figure}

The vector dot product was calculated for vectors of sizes $n_1 \in \{10^3, 10^4, 10^5, 10^6, 10^7, 10^8\}$. The average runtimes versus vector size for Python, Rust, using Rust from Python, and using a \texttt{Numba}-compiled version of the Python code  are shown in Figure~\ref{fig:dot}. The results illustrate a runtime improvement greater than 2 orders of magnitude between using Rust from Python over using Python alone. At larger vector sizes, the plot illustrates negligible differences between the average runtime performance between using Rust from Python, Rust, and using \texttt{Numba}. However, at smaller vector sizes, such as $n_1 =10^4$, the Rust implementation is faster than using Rust from Python and using \texttt{Numba}, potentially indicating runtime variability in these other methods at smaller vector sizes. 

Estimated overhead costs for the first call of using Rust from Python and the \texttt{Numba} JIT compiler as a function of the size of the vector are shown in Table~\ref{tab:dot_overhead}. The runtime overhead for the first call of using Rust from Python is approximately 3 orders of magnitude faster than the first call of using \texttt{Numba}. However, the estimated overhead for using Rust from Python increases slightly with the size of the vector, while the estimated overhead using \texttt{Numba} is relatively constant.

\begin{table}[htbp]
	\centering
	\caption{Estimated overhead costs (in seconds) for using Rust from Python and using the \texttt{Numba} Python JIT compiler for the vector dot product.}
	\begin{tabular}{|c|c|c|} 
		\hline
		Size of Vector & Rust from Python &  \texttt{Numba} \\ 
		\hline
		$10^3$ & $1.44\times 10^{-4}$& $3.17 \times 10^{-1}$\\
		$10^4$ & $1.44 \times 10^{-4}$& $3.17 \times 10^{-1}$\\
		$10^5$ & $1.49 \times 10^{-4}$& $3.18 \times 10^{-1}$\\
		$10^6$ & $1.58\times 10^{-4}$& $3.16 \times 10^{-1}$\\
		$10^7$ & $2.42 \times 10^{-4}$& $3.19 \times 10^{-1}$\\
		$10^8$ & $3.07 \times 10^{-4}$& $3.19 \times 10^{-1}$\\
		\hline
	\end{tabular}
	\label{tab:dot_overhead}
\end{table}

\subsection {Dense Matrix-Vector Product}
\label{sec:results:matvec}
The dense matrix-vector product was tested using matrices with an 1) increasing number of rows, 2) increasing number of columns, and 3) equal, increasing number of rows and columns of the following sizes:
\begin{itemize}
	\item $n_1 \in \{10^2, 10^3, 10^4, 10^5, 10^6\}$ with constant $n_2=10^2$;
	\item $n_2 \in \{10^2, 10^3, 10^4, 10^5, 10^6\}$ with constant $n_1=10^2$; and
	\item  $n_1, n_2 \in \{10^2,10^3, 10^4\}$.
\end{itemize}

\begin{figure}[ht!]
	\centering
	\begin{tabular}{cc}
		\includegraphics[height=0.275\textheight]{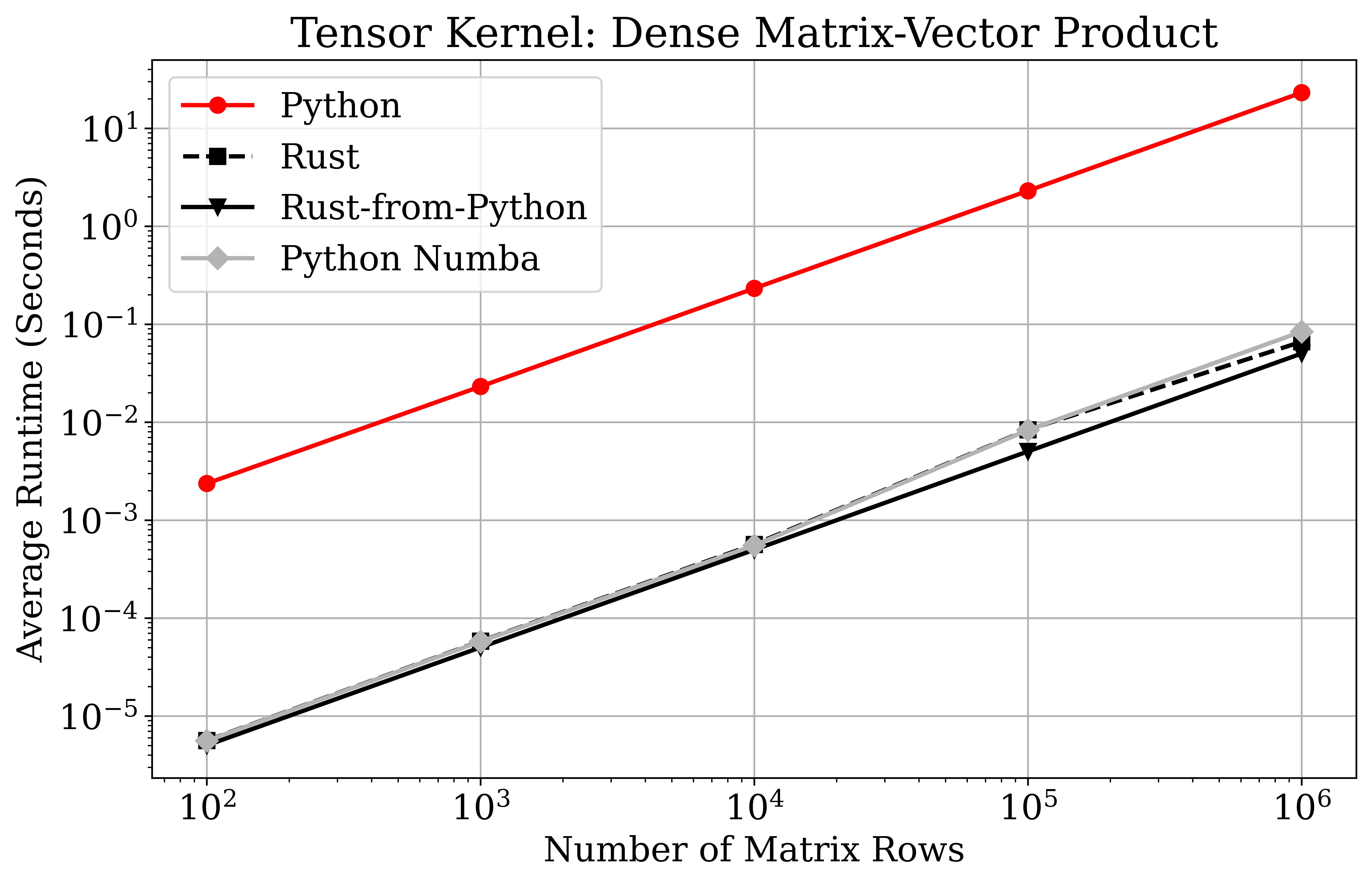}\\
		(a)\\
		\includegraphics[height=0.275\textheight]{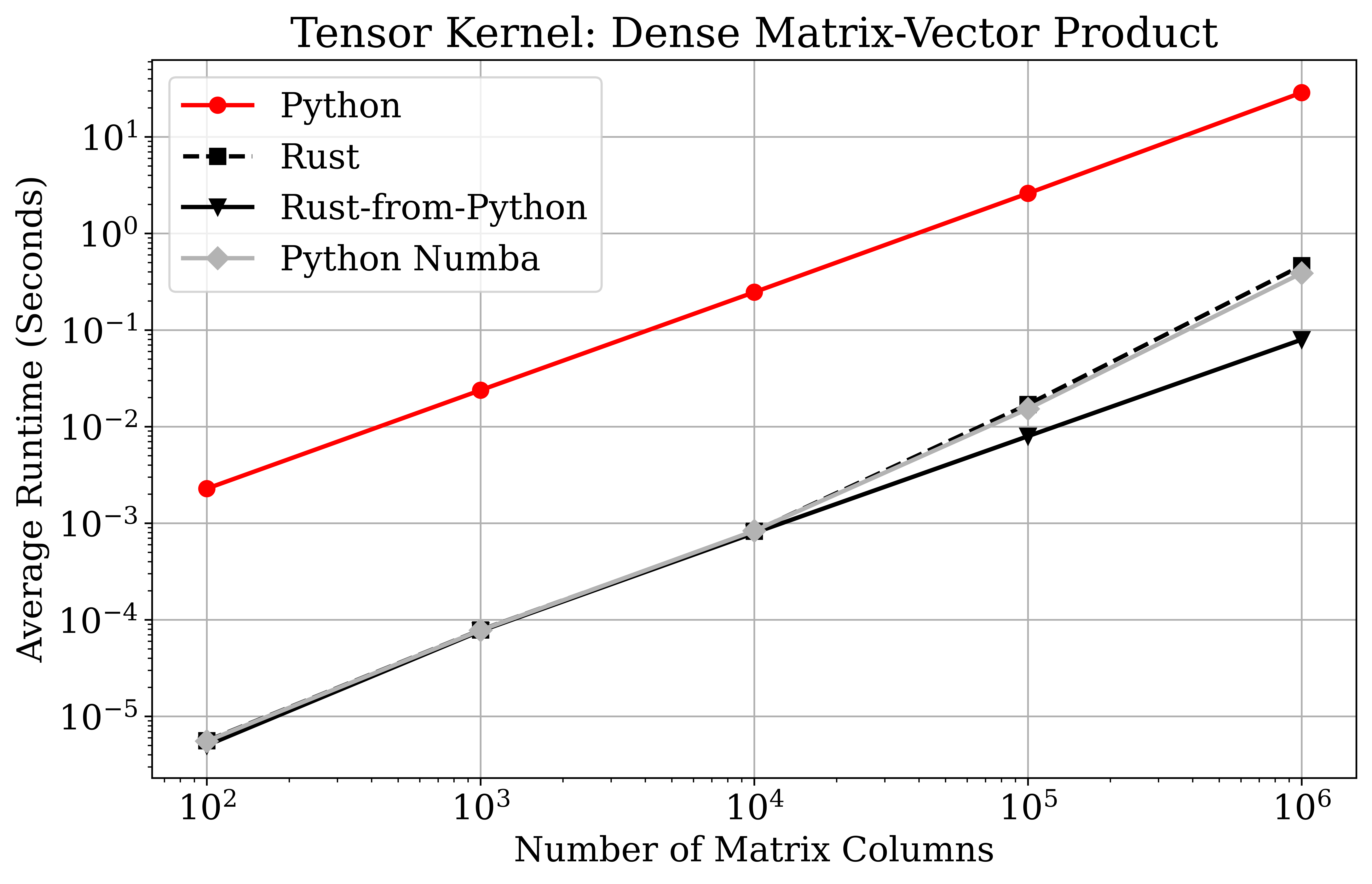}\\
		(b)\\
			\includegraphics[height=0.275\textheight]{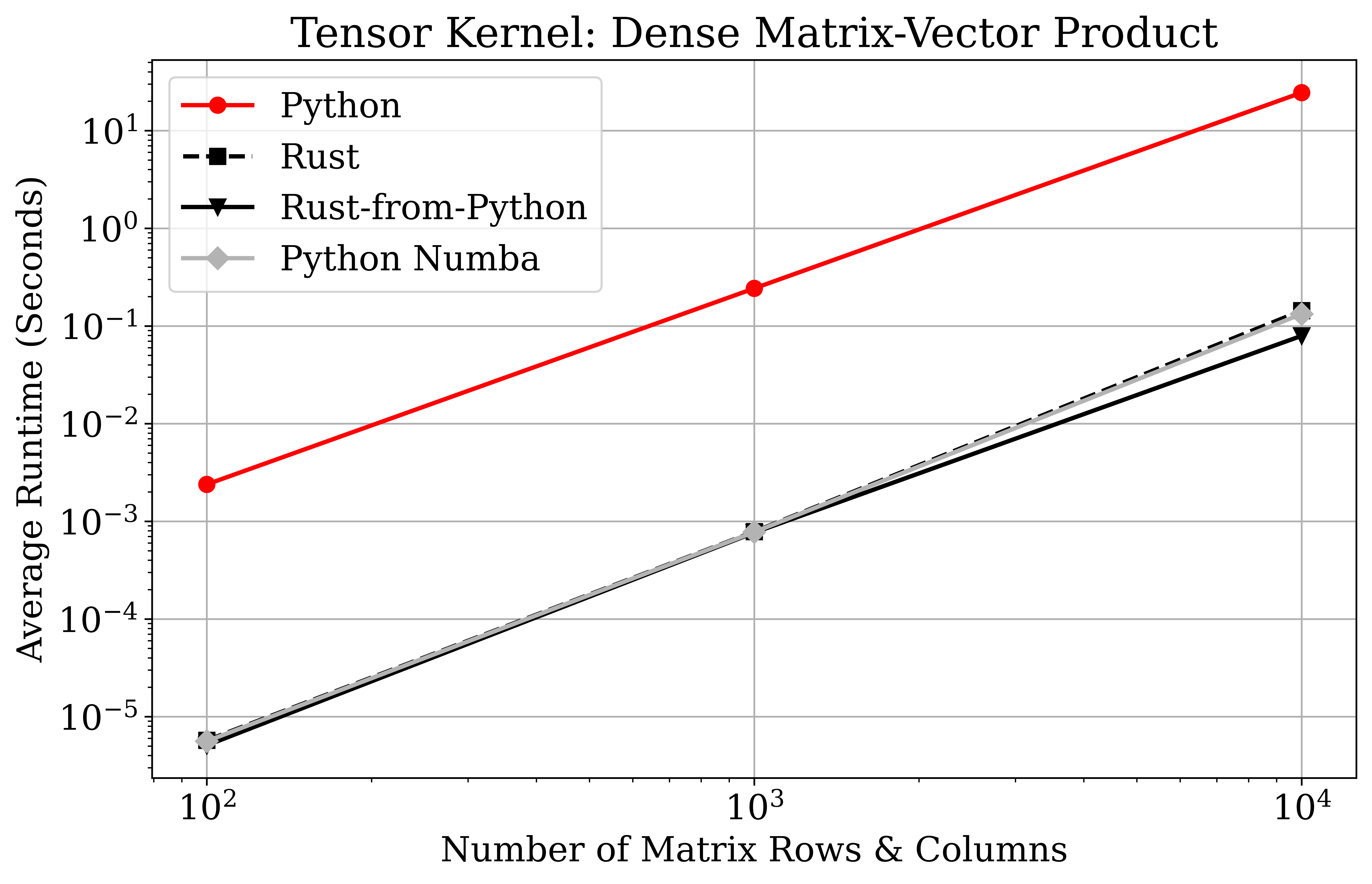}\\
		(c)
	\end{tabular}
	\caption{Average runtimes of dense matrix-vector product for (a) increasing number of columns, (b) increasing number of columns, and (c) equal, increasing number of rows and columns.}
	\label{fig:mvp}
\end{figure}

The average runtimes for Python, Rust, using Rust from Python, and using a \texttt{Numba}-compiled version of the Python code for the various cases are shown in Figure~\ref{fig:mvp}. At the largest matrix size, as demonstrated in all subplots, using Rust from Python presents a runtime performance improvement greater than 2 orders of magnitude. Note that the reasons for the increased average runtimes for Rust and \texttt{Numba} over using Rust from Python for the larger matrix sizes in all three experiments is unclear from our experiments. In future work, we plan to investigate the differences in more detail.

As in the results for the vector dot product discussed in Section~\ref{eq:vdp}, we estimated the overhead as a function of matrix size for the first call of using Rust through the FFI and JIT compiling with \texttt{Numba}. The results were calculated for matrices with equal numbers of rows and columns, with the total number of elements in the square matrices displayed in Table \ref{tab:mvp_overhead}. The results illustrate lower overhead from the first call of using Rust from Python in comparison to the first call when using \texttt{Numba}. However, the estimated overhead cost when using Rust from Python has greater variability as the size of the matrix increases in comparison to using \texttt{Numba}. 

\begin{table}[htbp]
	\centering
	\caption{Estimated overhead costs (in seconds) for using Rust from Python and using the \texttt{Numba} Python JIT compiler for the dense matrix-vector product.}
	\begin{tabular}{|c|c|c|} 
		\hline
		Number of Matrix Elements& Rust from Python &  \texttt{Numba} \\ 
		\hline
		$10^4$ & $1.50 \times 10^{-4}$& $3.58\times 10^{-1}$\\
		$10^6$ & $1.87\times 10^{-4}$& $3.58 \times 10^{-1}$\\
		$10^8$ & $6.41 \times 10^{-2}$& $3.06 \times 10^{-1}$\\
		\hline
	\end{tabular}
	\label{tab:mvp_overhead}
\end{table}

\subsection{Sparse Tensor Times Vector~(TTV) Product}
\label{sec:results:ttv}
The sparse tensor times vector product was tested using third-order tensors containing 1\% nonzero values (specified as \verb|density=0.01| when calling \verb|pyttb.sptenrand|) and equal sizes of dimensions, with
$n_1, n_2, n_3 \in \{100, 200, \dots, 1200\}$. In all experiments, the sparse tensors are multiplied along the second-dimension by a dense vector of random floating point values.

\begin{figure}[ht!]
	\centering
	\includegraphics[height=0.275\textheight]{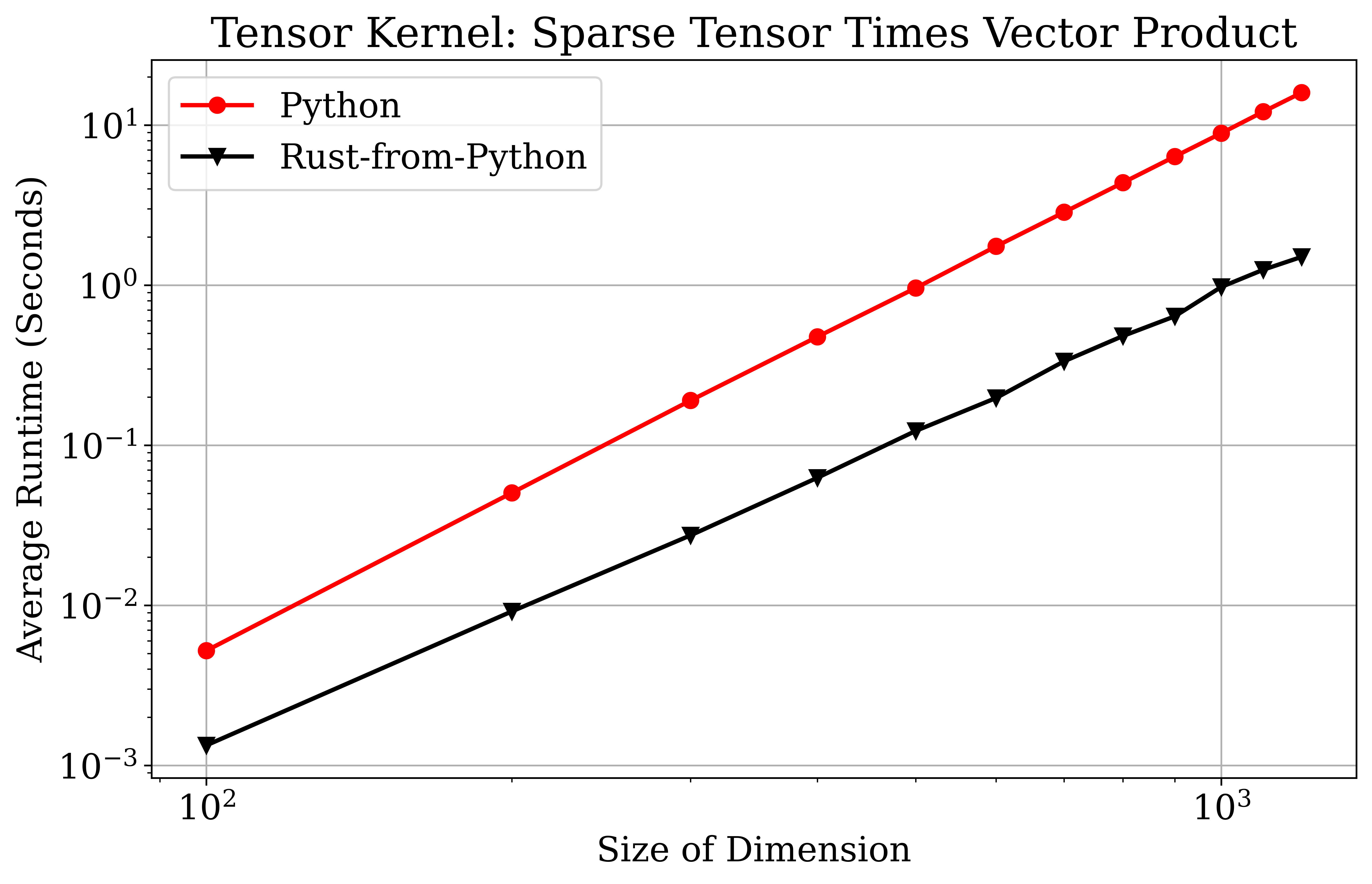}
	\caption{Average runtime of sparse tensor times vector product as a function of tensor dimension size.}
	\label{fig:ttv}
\end{figure}

The results shown in Figure \ref{fig:ttv} demonstrate an improved runtime performance of approximately 1 order of magnitude when using Rust from Python over using the Python implementation in \texttt{pyttb}. The plot illustrates that the difference between the runtime performance of using Rust from Python to Python increasing slightly with the size of the tensor. 
There are potentially many differences in the implementations leading to these runtime performance differences, such as memory layout, compiler optimizations (e.g., Rust iterators versus \texttt{NumPy} vectorized optimizations), and FFI boundary crossing overhead (Rust via \texttt{PyO3} versus \texttt{NumPy}, which uses C-based extension modules).	In future work, we plan to investigate the specific causes of these runtime improvements across these various factors.

\section{Conclusion and Future Work}
\label{sec:conc}
We demonstrated that using Rust from Python can result in improved runtime performance for several tensor kernels available in the Python Tensor Toolbox (\texttt{pyttb}). Our results indicate runtime improvements greater than 2 orders of magnitude for two simple tensor kernels---the vector dot product and the dense matrix-vector product---and runtime improvements of approximately 1 order of magnitude for a more advanced tensor kernel, the sparse tensor times vector (TTV) product. For the two former kernels, we also compared to versions of the Python implementations that were compiled into machine code using the just-in-time compiler from the \texttt{Numba} Python package, and results indicate comparable runtime performance improvements to that of using Rust from Python. Future work could consider experiments leveraging \texttt{Numba} on more advanced computation, like that of the sparse TTV kernel. There is a trade-off between lower development complexity versus higher one-time execution overhead cost in using just-in-time compilation in \texttt{Numba} that could also be explored in detail in future work. Other ideas for future work include investigation into the role of memory layout, compiler optimizations, using iterators over loop-based implementations, and increased computational complexity (e.g., as in the matricized tensor times Khatri-Rao product [MTTKRP]) when searching for opportunities that could lead to improved runtime performance. Lastly, although we focused here on runtime improvements on a single CPU core, we could also investigate the use of advanced computing architectures---e.g., GPUs (graphics processing units), TPUs (tensor processing units), etc.---and the use of on-node concurrency (via Rust or solutions provided in several Python packages) or distributed-memory computation to take advantage of multiple computing resources simultaneously.

    \section*{Acknowledgements} 
    We would like to express our sincere gratitude to Joshua B. Teves, Carolyn Mayer, Carlos Llosa, Rich Lehoucq, and Jeremy Myers for their invaluable feedback and suggestions during the preparation of this manuscript. Their insights greatly enhanced the quality of this work.
    
	\bibliographystyle{plain}
	\bibliography{ref}

\begin{thebibliography}{10}

\bibitem{TTB_MATLAB}
Brett~W. Bader and Tamara~G. Kolda.
\newblock {Tensor Toolbox for MATLAB (Version 3.6)}.
\newblock \url{https://www.tensortoolbox.org/}, 2023.

\bibitem{BaKo2025}
Grey Ballard and Tamara~G. Kolda.
\newblock {\em Tensor Decompositions for Data Science}.
\newblock Cambridge University Press, 2025.

\bibitem{Col2022}
Samuel Colvin.
\newblock {Pydantic V2 Plan}.
\newblock \url{https://pydantic.dev/articles/pydantic-v2}, 2022.
\newblock Accessed: 2025-07-02.

\bibitem{pyttb}
Daniel~M. Dunlavy, Nicholas~T. Johnson, et~al.
\newblock {pyttb: Python Tensor Toolbox (Version 1.8.2)}.
\newblock \url{https://github.com/sandialabs/pyttb}, January 2025.

\bibitem{2020NumPy-Array}
Charles~R. Harris, K.~Jarrod Millman, Stéfan~J van~der Walt, Ralf Gommers,
  Pauli Virtanen, David Cournapeau, Eric Wieser, Julian Taylor, Sebastian Berg,
  Nathaniel~J. Smith, Robert Kern, Matti Picus, Stephan Hoyer, Marten~H. van
  Kerkwijk, Matthew Brett, Allan Haldane, Jaime Fernández~del Río, Mark
  Wiebe, Pearu Peterson, Pierre Gérard-Marchant, Kevin Sheppard, Tyler Reddy,
  Warren Weckesser, Hameer Abbasi, Christoph Gohlke, and Travis~E. Oliphant.
\newblock Array programming with {NumPy}.
\newblock {\em Nature}, 585:357–362, 2020.

\bibitem{10.1007/978-3-031-85902-1_4}
Price~D. Johnson and Douglas~D. Hodson.
\newblock {PyO3}: Building {Python} extension modules in native {Rust} with
  performance and safety in mind.
\newblock In Douglas~D. Hodson, Michael~R. Grimaila, Hamid~R. Arabnia, Leonidas
  Deligiannidis, and Torrey~J. Wagner, editors, {\em Scientific Computing and
  Bioinformatics and Computational Biology}, pages 23--30, Cham, 2025. Springer
  Nature Switzerland.

\bibitem{Kh2018}
Boris~N. Khoromskij.
\newblock {\em Tensor Numerical Methods in Scientific Computing}.
\newblock De Gruyter, Berlin, Boston, 2018.

\bibitem{Rust_Book}
Steve Klabnik, Carol Nichols, et~al.
\newblock {\em The Rust Programming Language, 2nd Edition}.
\newblock No Starch Press, 2022.

\bibitem{TaHuNiMa2025}
Taichi Kosugi, Xinchi Huang, Hirofumi Nishi, and Yu-ichiro Matsushita.
\newblock Tensor-decomposition technique for qubit encoding of maximal-fidelity
  lorentzian orbitals in real-space quantum chemistry.
\newblock {\em Phys. Rev. A}, 111:052615, May 2025.

\bibitem{10.1145/2833157.2833162}
Siu~Kwan Lam, Antoine Pitrou, and Stanley Seibert.
\newblock Numba: a {LLVM}-based {Python} {JIT} compiler.
\newblock In {\em Proceedings of the Second Workshop on the LLVM Compiler
  Infrastructure in HPC}, LLVM '15, New York, NY, USA, 2015. Association for
  Computing Machinery.

\bibitem{Lock18}
Eric~F Lock.
\newblock Tensor-on-tensor regression.
\newblock {\em Technometrics}, 60(4):480--492, 2018.

\bibitem{MaYa2022}
Linjian Ma and Chao Yang.
\newblock Low rank approximation in simulations of quantum algorithms.
\newblock {\em Journal of Computational Science}, 59:101561, 2022.

\bibitem{ndarray_rust}
{Ndarray Developers}.
\newblock Ndarray: an {N}-dimensional array for general elements and numerics
  (version 0.16.1).
\newblock \url{https://github.com/rust-ndarray/ndarray}, 2025.

\bibitem{tiktoken}
OpenAI.
\newblock tiktoken.
\newblock \url{https://github.com/openai/tiktoken}, 2023.

\bibitem{PaKoChOl2024}
Yannis Panagakis, Jean Kossaifi, Grigorios~G Chrysos, James Oldfield, Taylor
  Patti, Mihalis~A Nicolaou, Anima Anandkumar, and Stefanos Zafeiriou.
\newblock Tensor methods in deep learning.
\newblock In {\em Signal Processing and Machine Learning Theory}, pages
  1009--1048. Academic Press, 2024.

\bibitem{Ped2023}
Ulrik~Thyge Pederson.
\newblock Better together four examples of how {Rust} makes {Python} better.
\newblock
  \url{https://towardsai.net/p/l/better-together-four-examples-of-how-rust-makes-python-better},
  2023.
\newblock Accessed: 2025-07-02.

\bibitem{PeStCa2026}
A.~Pellegrino, H.~Stein, and N.A. Cayco-Gajic.
\newblock Dimensionality reduction beyond neural subspaces with slice tensor
  component analysis.
\newblock {\em Nature Neuroscience}, 27:1199--1210, 2024.

\bibitem{PeDu2014}
Matthew~G. Peterson and Daniel~M. Dunlavy.
\newblock {Tensor Toolbox: Wrapping to Python using SWIG}, December 2014.
\newblock Technical Report Number SAND2015-3829O.

\bibitem{maturin}
{PyO3 Project and Contributors}.
\newblock {Maturin}.
\newblock \url{https://github.com/PyO3/maturin}, 2025.

\bibitem{PyO3_Project_and_Contributors_PyO3}
{PyO3 Project and Contributors}.
\newblock {PyO3 (Version 0.25.0)}.
\newblock \url{https://github.com/PyO3/pyo3}, 2025.

\bibitem{freeThreadingPython}
{Python Software Foundation}.
\newblock Python experimental support for free threading.
\newblock \url{https://docs.python.org/3/howto/free-threading-python.html}.
\newblock Accessed: 2025-07-02.

\bibitem{pythoncapi}
{Python Software Foundation}.
\newblock {Python C API Reference Manual}.
\newblock \url{https://docs.python.org/3/c-api/index.html}, 2025.
\newblock Version 3.x (refer to specific version if applicable).

\bibitem{Rao2023}
Prashanth Rao.
\newblock How {Rust} is supercharging {Python} from the ground up.
\newblock \url{https://thedataquarry.com/blog/rust-is-supercharging-python},
  2023.
\newblock Accessed: 2025-07-02.

\bibitem{TOKCAN2026110191}
Neriman Tokcan, Shakir~Showkat Sofi, Van~Tien Pham, Clémence Prévost, Sofiane
  Kharbech, Baptiste Magnier, Thanh~Phuong Nguyen, Yassine Zniyed, and Lieven
  {De Lathauwer}.
\newblock Tensor decompositions for signal processing: Theory, advances, and
  applications.
\newblock {\em Signal Processing}, 238:110191, 2026.

\bibitem{van1995python}
Guido Van~Rossum and Fred~L Drake~Jr.
\newblock {\em Python reference manual}.
\newblock Centrum voor Wiskunde en Informatica Amsterdam, 1995.

\end{thebibliography}
	
\end{document}